\documentclass[twocolumn,showpacs,prl,floatfix,amsmath,amssymb,notitlepage,superscriptaddress]{revtex4-1}
\usepackage{epsfig}
\usepackage{subfigure}
\usepackage{amsmath}
\usepackage{color}
\usepackage{amssymb}
\usepackage{setspace}
\usepackage{graphicx}% Include figure files
\usepackage{dcolumn}% Align table columns on decimal point
\usepackage{bm}% bold math
\usepackage{times}
\usepackage{enumerate}
\usepackage{footnote}
\input epsf

\graphicspath{{figures/}}

\begin{document}

\author{Per Sebastian Skardal}
\email{persebastian.skardal@trincoll.edu} 
\affiliation{Department of Mathematics, Trinity College, Hartford, CT 06106, USA}

\author{Alex Arenas}
\affiliation{Departament d'Enginyeria Informatica i Matem\'{a}tiques, Universitat Rovira i Virgili, 43007 Tarragona, Spain}

\title{Abrupt Desynchronization and Extensive Multistability in Globally Coupled Oscillator Simplices}
%\date{\today}

\begin{abstract}
Collective behavior in large ensembles of dynamical units with non-pairwise interactions may play an important role in several systems ranging from brain function to social networks. Despite recent work pointing to simplicial structure, i.e., higher-order interactions between three or more units at a time, their dynamical characteristics remain poorly understood.  Here we present an analysis of the collective dynamics of such a simplicial system, namely coupled phase oscillators with three-way interactions. The simplicial structure gives rise to a number of novel phenomena, most notably a continuum of abrupt desynchronization transitions with no abrupt synchronization transition counterpart, as well as extensive multistability whereby infinitely many stable partially synchronized states exist. Our analysis sheds light on the complexity that can arise in physical systems with simplicial interactions like the human brain and the role that simplicial interactions play in storing information.
\end{abstract}


\maketitle

Research into the macroscopic dynamics of large ensembles of coupled oscillators have extended our understanding of natural and engineered systems ranging from cell cycles to power grids~\cite{Strogatz2003,Pikovsky2003,Prindle2012Nature,Rohden2012PRL,Skardal2015SciAdv}. However, with few exceptions (including \cite{Tanaka2011PRL,Komarov2015PRE,Bick2016Chaos}), little attention has been paid to the synchronization dynamics of coupled oscillator systems where interactions are not pair-wise, but rather $n$-way, with $n\ge3$. Such interactions are called ``simplicial'', where an $n$-simplex represents an interaction between $n+1$ units, so $2$-simplices describe three-way interactions, etc~\cite{Salnikov2019EJP}. Recent advances suggest that simplicial interactions may be vital in general oscillator systems~\cite{Ashwin2016PhysD,Gin2018,Gin2019} and may play an important role in brain dynamics~\cite{Petri2014Interface,Giusti2016JCN,Sizemore2018JCN} and other complex systems phenomena such as, the dynamics of collaborations~\cite{Patania2017EPJDS} or social contagion~\cite{Iacopini2019}. In particular, interactions in 2-simplices (named holes or cavities) are important because they can describe correlations in neuronal spiking activity (that can be mapped to phase oscillators~\cite{Politi2015}) in the brain~\cite{Reimann2017} providing a missing link between structure and function. In fact, coupled oscillator systems that display clustering and multi-branch entrainment have been shown to be useful models for memory and information storage~\cite{Daido1996PRL,Hoppensteadt1999PRL,Fell2011Nature,Hipp2012Nature,Ashwin2007SIAM,Skardal2011PRE,Komarov2013PRL}. Despite these findings, the general collective dynamics of coupled oscillator simplices and their utility in storing information are poorly understood.

In this work we study large coupled oscillator simplicial complexes, considering the \textcolor{black}{impact} of $2$-simplices, i.e., three-way interactions, \textcolor{black}{on collective behavior. Specifically, we consider the 2- and 1-simplex multilayer system given by}
\textcolor{black}{\begin{align}
\dot{\theta}_i &= \omega_i + \frac{K}{N^2}\sum_{j=1}^N\sum_{k=1}^N\sin\left(\theta_{j}+\theta_{k}-2\theta_i\right),\label{eq:01}\\
\phi_i&=\nu_i +\frac{\kappa}{N}\sum_{j=1}^N\sin(\phi_j-\phi_i) + d\sin(\theta_i-\phi_i).\label{eq:02}
\end{align}}
\textcolor{black}{The dynamics in the $\theta$-layer are the natural generalization of the classical Kuramoto model~\cite{Kuramoto1984} with $2$-simplex interactions (namely, coupling is sinusoidal and diffusive),
where $\theta_i$ represents the phase of oscillator $i$ with $i=1,\dots,N$, $\omega_i$ is its natural frequency which is assumed to be drawn from the distribution $g(\omega)$, and $K$ is the global coupling strength. The $\theta$-oscillators then drive a population of $\phi$-oscillators in another layer, for which a one-to-one correspondence exists following a multiplex structure~\cite{DeDomenico2013PRX}. Specifically, the $\phi$-oscillators evolve subject to the classical Kuramoto model (with natural frequencies $\nu_i$ and coupling strength $\kappa$), but with an additional driving term with strength $d$.
}
While numerical investigations of systems with non-pairwise interactions have uncovered multistability and chaos~\cite{Tanaka2011PRL,Komarov2015PRE,Bick2016Chaos}, few analytical results exist and their collective behavior remains largely unexplored. Here we focus on large systems and obtain an analytical description of the macroscopic dynamics using a partial dimensionality reduction obtained via a variation of the Ott-Antonsen ansatz~\cite{Ott2008Chaos,Ott2009Chaos}. In particular, the \textcolor{black}{2-simplex} macroscopic dynamics are captured by a combination of two order parameters that capture the degree of synchronization and asymmetry as oscillators organize into two distinct synchronized clusters. Here, clustering refers to multiple phase-locked groups~\cite{Daido1996PRL,Skardal2011PRE} at different locations on the circle rather than distinct groups of identically synchronized oscillators~\cite{Pecora2014NatComm}. We uncover a novel phenomenon where a continuum of abrupt desynchronization transitions emerge, each at a different critical coupling strength, depending on the asymmetry of the system. Interestingly, no complementary abrupt synchronization transitions occur~\cite{GomezGardenes2011PRL,Skardal2014PRE}. This continuum stems from an extensive multistability whereby, for sufficiently strong coupling, an infinite number of distinct partially synchronized states are stable in addition to the incoherent state, which is stable for all finite coupling strengths. This multistability indicates the capability of storing a wide array of possible information as different oscillator arrangements. Serving as a minimal model for memory storage, the system captures the critical properties of easily transitioning from an information storage state (i.e., synchronized) to the resting state (i.e., incoherent)~\cite{Deco2013Trends} via abrupt desynchronization. The system then may return to another information storage state with an appropriately chosen perturbation. \textcolor{black}{Moreover, the 1-simplex layer can similarly store information provided that the driving strength from the 2-simplex layer is sufficiently strong.} The rich nonlinear dynamics that emerge in this relatively simple extension of pair-wise coupling to three-way coupling highlights the complexity that may arise via simplicial interactions in systems like the human brain and the implications of these behaviors on information storage.

\textcolor{black}{Noting that the dynamics of the 1-simplex layer in Eq.~(\ref{eq:02}) are driven by the dynamics of the 2-simplex layer in Eq.~(\ref{eq:01}), we focus first on the $\theta$-layer} our analysis by introducing the generalized order parameters $z_q=\frac{1}{N}\sum_{j=1}^Ne^{qi\theta_j}$ for $q=1$ and $2$. Note that $z_1$ is the classical Kuramoto order parameter while $z_2$ typically measures clustering, which we will see in this system. Using the polar decompositions $z_q=r_qe^{i\psi_q}$ we rewrite Eq.~(\ref{eq:01}) as
\begin{align}
\dot{\theta}_i=\omega_i + K r_1^2\sin[2(\psi_1-\theta_i)].\label{eq:03}
\end{align}
We then consider the continuum limit $N\to\infty$ where the state of the system can be described by a density function $f(\theta,\omega,t)$ which describes the density of oscillator with phase between $\theta$ and $\theta+\delta\theta$ and natural frequency between $\omega$ and $\omega+\delta\omega$ at time $t$. Because the number of oscillators in the system is conserved $f$ must satisfy the continuity equation $0=\partial_t f + \partial_\theta(f\dot{\theta})$. Moreover, because each oscillator's natural frequency is fixed and drawn from $g(\omega)$ the density function $f(\theta,\omega,t)$ may be expanded into a Fourier series of the form $f(\theta,\omega,t)=\frac{g(\omega)}{2\pi}\left[1+\sum_{n=1}^\infty \hat{f}_n(\omega,t)e^{in\theta}+c.c.\right]$, where $\hat{f}_n(\omega,t)$ is the $n^{\text{th}}$ Fourier coefficient and $c.c.$ represents the complex conjugate of the previous sum. 

We then consider the symmetric and asymmetric parts $f_s(\theta,\omega,t)$ and $f_a(\theta,\omega,t)$, respectively, of $f(\theta,\omega,t)$ which satisfy $f(\theta,\omega,t)=f_s(\theta,\omega,t)+f_a(\theta,\omega,t)$ with symmetries $f_s(\theta+\pi,\omega,t)=f_s(\theta,\omega,t)$ and $f_a(\theta+\pi,\omega,t)=-f_a(\theta,\omega,t)$. Note that the linearity of the continuity equation implies that if both $f_s$ and $f_a$ are solutions, then so is $f$. While the asymmetric part $f_a$ does not allow for dimensionality reduction, the symmetric part $f_s$ does. Noting that the Fourier series of $f_s$ is given by the even terms of the Fourier series of $f$, i.e., $f_s(\theta,\omega,t)=\frac{g(\omega)}{2\pi}\left[1+\sum_{m=1}^\infty \hat{f}_{2m}(\omega,t)e^{2im\theta}+c.c.\right]$, we make the ansatz that each even Fourier coefficient decays geometrically, i.e., $\hat{f}_{2m}(\omega,t)=a^m(\omega,t)$~\cite{Ott2008Chaos,Ott2009Chaos}. Inserting this and Eq.~(\ref{eq:02}) into the continuity equation, we find that each subspace spanned by even terms $e^{2im\theta}$ collapse onto the same low-dimensional manifold characterized by the condition
\begin{align}
\partial_t a=-2i\omega a+K\left(z_{1}^{*2}-z_{1}^2 a^2\right).\label{eq:04}
\end{align}

Equation~(\ref{eq:04}) describes the evolution of the complex function $a(\omega,t)$, and thereby $f_s$, and depends on the order parameter $z_1$. Moreover, $a(\omega,t)$ can be linked to the order parameter $z_2$ as follows. First, note that in the limit $N\to\infty$ we have that $z_2=\iint f_s(\theta,\omega,t)e^{2i\theta}d\theta d\omega$, and after inserting the Fourier series for $f_s$ this reduces to $z_2=\int g(\omega)a^*(\omega,t)d\omega$. To further simplify the relationship we make the assumption that the frequency distribution $g(\omega)$ is Lorentzian with mean $\omega_0$ and spread $\Delta$, i.e., $g(\omega)=\Delta/\{\pi\left[(\omega-\omega_0)^2+\Delta^2\right]\}$, which has two simple poles in the complex plane at $\omega = \omega_0\pm i\Delta$. The integral can then be evaluated using Cauchy's Residue Theorem~\cite{Ablowitz2003} by closing the integral contour with a semicircle of infinite radius in the lower-half plane and evaluating at the enclosed pole, yielding $z_2=a^*(\omega_0-i\Delta,t)$. We then evaluate Eq.~(\ref{eq:04}) at $\omega=\omega_0-i\Delta$ to obtain
\begin{align}
\dot{z}_2=2i\omega_0z_2-2\Delta z_2+K\left(z_{1}^2-z_{1}^{*2}z_2^2\right).\label{eq:05}
\end{align}
Next we introduce the rescaled parameters $\tilde{K}=K/\Delta$ and $\tilde{\omega}_0=\omega_0/\Delta$ with rescaled time $\tilde{t}=\Delta t$, then use polar decompositions, yielding (where we have dropped the $\sim$-notation for convenience)
\begin{align}
\dot{r}_2&=-2r_2+Kr_{1}^2(1-r_2^2)\cos(2\psi_{1}-\psi_2),\label{eq:06}\\
\dot{\psi}_2&=2\omega_0+Kr_{1}^2\frac{1+r_2^2}{r_2}\sin(2\psi_{1}-\psi_2).\label{eq:07}
\end{align}
Equations~(\ref{eq:06}) and (\ref{eq:07}) describe the dynamics of the even part $f_s$, which falls onto a low dimensional manifold similar to the Ott-Antonsen manifold and describes the evolution of $z_2$. However, these equations do not capture the asymmetric part of the dynamics, and moreover they depends on the asymmetric part via $z_2$'s dependence on $z_1$. As we will see, this reflects the system's dependence on asymmetry in oscillator arrangements between two clusters. 

To close the dynamics we apply a self-consistency analysis to characterize the order parameter $z_1$. We first note that, by entering the rotating frame $\theta\mapsto\theta+\omega_0t$ we set $\omega_0=0$ so that $\dot{\psi}_1=\dot{\psi}_2=0$. Moreover, by rotating initial conditions $\theta(0)\mapsto\theta(0)+\psi_1(0)$ we set $\psi_1=\psi_2=0$. Equation~(\ref{eq:02}) then implies that oscillators that become phase-locked satisfy $|\omega_i|\le Kr_1^2$, in which case they relax to one of the two stable fixed points $\theta_i=\theta^*(\omega_i)$ or $\theta^*(\omega_i)+\pi$, where $\theta^*(\omega)=\arcsin(\omega/Kr_1^2)/2$. These two fixed points correspond to the two clusters that the phase-locked oscillators organize into. Specifically, phase-locked oscillators starting near $\theta=0$ or $\pi$ will end up at the fixed points $\theta^*(\omega)$ or $\theta^*(\omega)+\pi$, respectively. The phase-locked population is described by the density function
\begin{align}
f_{\text{locked}}(\theta,\omega)=\eta\delta(\theta-\theta^*(\omega)) + (1-\eta)\delta(\theta-\theta^*(\omega)-\pi),\label{eq:08}
\end{align}
where the asymmetry parameter $\eta$ describes the fraction of phase-locked oscillators in the $\theta=0$ cluster. On the other hand, oscillators satisfying $|\omega_i|>Kr_1^2$ drift for all time and relax to the stationary distribution
\begin{align}
f_{\text{drift}}(\theta,\omega) = \frac{\sqrt{\omega^2-K^2r_1^4}}{2\pi\left[\omega+Kr_1^2\sin(2\psi_1-2\theta)\right]}.\label{eq:09}
\end{align}
Next, the order parameter $z_1$ is given by the integral $z_1=\iint f(\theta,\omega,t)e^{i\theta}d\theta d\omega$, which after inserting the density $f$ as defined by Eqs.~(\ref{eq:08}) and (\ref{eq:09}) reduces to 
\begin{align}
r_1=(2\eta-1)\int_{-Kr_1^2}^{Kr_1^2}\sqrt{\frac{1+\sqrt{1-(\omega/Kr_1^2)^2}}{2}}g(\omega)d\omega,\label{eq:10}
\end{align}
where  the contribution from the drifting oscillators vanishes due to the symmetry of $f_{\text{drift}}$. Returning to $r_2$, Eq.~(\ref{eq:06}) implies that at steady state we have
\begin{align}
r_2 = \frac{-1 + \sqrt{1 + K^2r_1^4}}{Kr_1^2}.\label{eq:11}
\end{align}
Thus, the macroscopic steady-state is described by Eqs.~(\ref{eq:10}) and (\ref{eq:11}).

\begin{figure}[t]
\centering
\epsfig{file =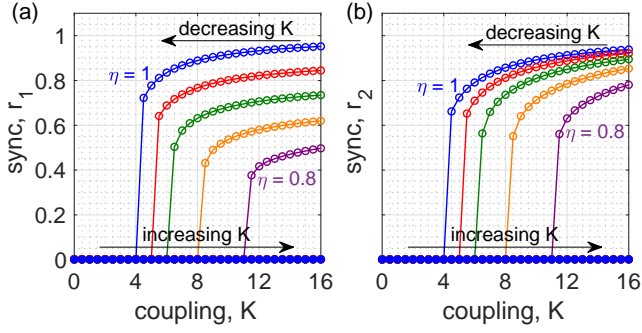, clip =,width=1.0\linewidth }
\caption{{\it Synchronization profiles.} The order parameters (a) $r_1$ and (b) $r_2$ as a function of the coupling strength $K$ for various asymmetry values $\eta$. Blue, red, green, orange, and purple circles represent results obtained from direct simulations of Eq.~(\ref{eq:02}) with $N=10^5$ oscillators for $\eta=1$, $0.95$, $0.9$, $0.85$, and $0.8$, respectively.} \label{fig1}
\end{figure}

\begin{figure}[t]
\centering
\epsfig{file =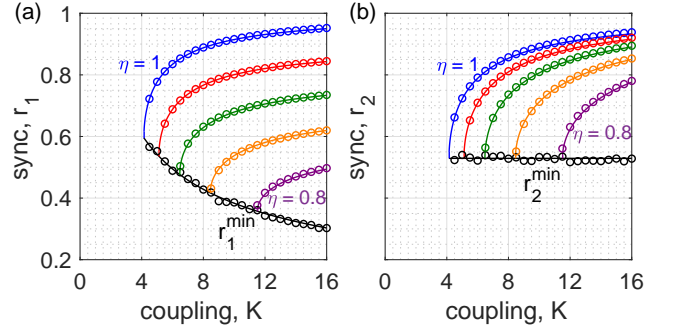, clip =,width=1.0\linewidth }
\caption{{\it Synchronization branches.} For order parameters (a) $r_1$ and (b) $r_2$, the synchronization branches predicted by Eqs.~(\ref{eq:10}) and (\ref{eq:11}) plotted as solid curves for asymmetries $\eta=1$, $0.95$, $0.9$, $0.85$, and $0.8$ as well as the minimum branches [predictions for which are given in Eq.~(\ref{eq:14})]. Results obtained from direct simulations with $N=10^5$ oscillators are plotted in blue, red, green, orange, and purple circles for each value of $\eta$ and black circles represent minimum branches.} \label{fig2}
\end{figure}

Interpreting these analytical results in the context of numerical simulations allows us to understand novel phenomena that occur in the dynamics of Eq.~(\ref{eq:01}). Beginning with simulations of a system with $N=10^5$ oscillators whose natural frequencies are Lorentzian with $\omega_0=0$ and $\Delta=1$, we consider initial conditions of varying asymmetry, setting initial phases to $\theta_i(0)=0$ and $\pi$ with probabilities $\eta$ and $1-\eta$, respectively. We then begin simulations at $K=16$ and after reaching steady-state slowly decrease $K$ to zero, then restore it slowly to $16$. In Figs.~\ref{fig1}(a) and (b) we plot the resulting values for the order parameters $r_1$ and $r_2$, respectively, for $\eta=1$ (blue, top), $0.95$ (red), $0.9$ (green), $0.85$ (orange), and $0.8$ (purple, bottom). Arrows indicate the direction of $K$ and results corresponding to decreasing and increasing $K$ are plotted as open and filled circles, respectively. As $K$ is initially decreased solutions traverse different partially synchronized states that are determined by $\eta$, until each branch undergoes a discontinuous jump to the incoherent state at different critical coupling strengths in abrupt desynchronization transitions. Importantly, this highlights both a rich extensive multistability (here five different branches are shown, but in the thermodynamic limit an infinite number of such states exist) and a continuum of abrupt desynchronization transitions at different coupling strengths. Partially synchronized branches are characterized by the asymmetry parameter $\eta$, indicating that these complex dynamics arise from different allocations of phase-locked oscillators in the two clusters at $\theta=0$ and $\pi$. Next, as $K$ is restored to its initial value of $16$ no spontaneous transitions back to synchronization occur, indicating no abrupt synchronization to complement the abrupt desynchronization transitions. In Figs.~\ref{fig2}(a) and (b) we show that our theory captures these dynamics, plotting in solid curves the theoretical predictions of $r_1$ and $r_2$, respectively, given by Eqs.~(\ref{eq:10}) and (\ref{eq:11}). We use the same values of $\eta$ as in Fig.~\ref{fig1} and overlay the results from simulations in open circles, noting excellent agreement.

\begin{figure}[t]
\centering
\epsfig{file =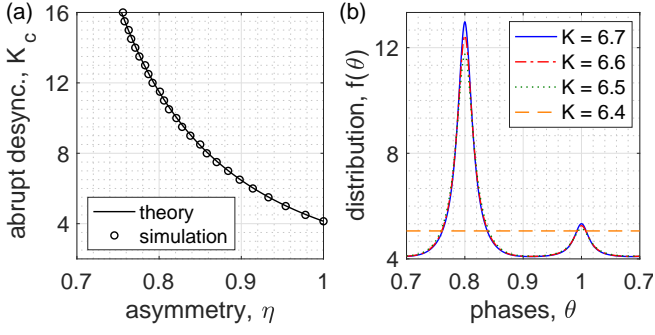, clip =,width=1.0\linewidth }
\caption{{\it Abrupt desynchronization transition.} (a) The critical coupling strength $K_c$ at which the abrupt desynchronization transition occurs as a function of the asymmetry $\eta$. The solid curve represents the theoretical prediction given by Eq.~(\ref{eq:14}) and black circles represent observations from direct simulation with $N=10^5$ oscillators. (b) Illustration of the distribution $f(\theta)$ of phases as the system passes though the abrupt desynchronization transition. Here $\eta=0.9$ with $K_c\approx6.47$. Distributions for $K=6.7$, $6.6$, $6.5$, and $6.4$ are plotted in solid blue, dot-dashed red, dotted green, and dashed orange curves.} \label{fig3}
\end{figure}

To better understand the nature of the abrupt desynchronization and multistability phenomena, we investigate the minimal partially synchronized states, i.e., minimal non-zero values $r_{1}^{\text{min}}$ and $r_{2}^{\text{min}}$ allowable for different coupling strengths. We first plot numerical results for $r_{1}^{\text{min}}$ and $r_{2}^{\text{min}}$ in Figs.~\ref{fig2}(a) and (b) using black circles. Interestingly, while $r_1^{\text{min}}$ appears to decay monotonically with $K$, $r_2^{\text{min}}$ remains roughly constant. In fact, the minimal branch $r_2^\text{min}$ is constant and can be leveraged to analytically describe the branches $r_1^{\text{min}}(K)$ and $r_2^{\text{min}}(K)$ as well as the corresponding asymmetry value $\eta_{\text{min}}(K)$ and critical coupling strengths $K_c(\eta)$ where the abrupt desynchronization occurs. We proceed by inverting Eq.~(\ref{eq:11}), obtaining $Kr_1^2=2r_2/(1-r_2^2)$, which can be inserted into Eq.~(\ref{eq:10}), yielding
\begin{align}
&\sqrt{\frac{2 r_2}{1-r_2^2}}=\sqrt{K}(2\eta-1)\nonumber\\
&~~\times\int_{-2 r_2/(1-r_2^2)}^{2 r_2/(1-r_2^2)}\sqrt{\frac{1+\sqrt{1-\left[\omega(1-r_2^2)/2r_2\right]^2}}{2}}g(\omega)d\omega.\label{eq:12}
\end{align}
While Eq.~(\ref{eq:12}) appears more complicated than Eq.~(\ref{eq:10}), we note that the coupling strength $K$ has been entirely scaled out of the integral, appearing outside with $(2\eta-1)$. We therefore conclude that if the quantities $\sqrt{K}$ and $2\eta-1$ cancel one another out, i.e., $\sqrt{K}(2\eta-1)$ is constant, it follows that the solution $r_2$ in Eq.~(\ref{eq:12}) is independent of $K$. We therefore propose the ansatz $\sqrt{K}(2\eta-1)=\text{const.}$ and use the initial condition $\eta_{\text{min}}(K_c(1))=1$, where $K_c(1)$ denotes the very first coupling strength where a synchronized state is possible with $\eta=1$, yielding
\begin{align}
\eta_{\text{min}}(K)=\frac{\sqrt{K_c(1)}}{2\sqrt{K}}+\frac{1}{2}.\label{eq:13}
\end{align}
Equation.~(\ref{eq:13}) implies that along the minimum branch we have that $\sqrt{K}(2\eta-1)=\sqrt{K_c(1)}\approx2.034$, which can be used in Eq.~(\ref{eq:12}) to compute the minimum branch of $r_2$, and in turn $r_1$ via Eq.~(\ref{eq:11}), yielding
\begin{align}
r_1^{\text{min}}(K)\approx\frac{1.2120}{\sqrt{K}}\hskip2ex\text{and}\hskip2exr_2^{\text{min}}(K)\approx 0.5290.\label{eq:14}
\end{align}
In Figs.~\ref{fig2}(a) and (b) we plot these minimum branches in solid black curves, which agree with the simulation results. Lastly, by inverting Eq.~(\ref{eq:13}) we find the critical coupling strength $K_c$ as a function of $\eta$ where the abrupt desynchronization transition occurs, namely
\begin{align}
K_c(\eta) \approx\frac{4.137}{(2\eta-1)^2}.\label{eq:14}
\end{align} 
In Fig.~\ref{fig3}(a) we plot the theoretical prediction of the abrupt desynchronization point $K_c(\eta)$ as a solid curve vs observations from direct simulations as black circles, noting excellent agreement. %We emphasize that above this curve we observe extensive multistability and below the curve only the incoherent state is stable.

\textcolor{black}{The dynamics in the 2-simplex layer} serves as a minimal model for memory and information storage, capturing a number of critical properties. Each distinct synchronized state corresponds to a specific piece of information, differentiated by the clustering arrangement of the oscillators. Moreover, synchronized state the system can quickly and easily transition to the resting state described by incoherence via the abrupt desynchronized transition. The microscopic properties of this abrupt desynchronization transition \textcolor{black}{are} illustrated in Fig.~\ref{fig3}(b), where for $\eta=0.9$ the distribution $f(\theta)$ of phases is plotted as $K$ is decreased slowly through the critical value of $K_c\approx6.47$. Before the transition the distribution is asymmetrically clustered about $\theta=0$ and $\pi$ and changes slowly until at $K=K_c$ all information is lost as the distribution becomes uniform\textcolor{black}{, representing a resting state.} %Once the incoherent resting state is reached, the system then may return to any synchronized (i.e., information storage) state via an appropriately designed perturbation that pins different oscillator to the $\theta=0$ or $\pi$ states, depending on the desired target state.

\begin{figure}[t]
\centering
\epsfig{file =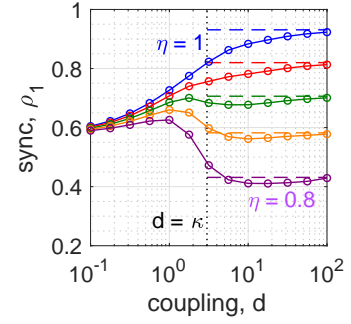, clip =,width=0.55\linewidth }
\caption{ \textcolor{black}{ {\it 1-simplex layer dynamics.} The order parameter $\rho_1$ describing synchronization of $\phi$-oscillators in the 1-simplex layer as a function of the driving strength $d$. Connected circles represent numerical simulations with $10^5$ oscillators. 2-simplex and 1-simplex coupling are $k=12$ and $\kappa = 3$, and asymmetry values $\eta=1$, $0.95$, $0.9$, $0.85$, and $0.8$ are colored blue, red, green, orange, and purple, respectively. dashed horizontal lines denote the values of $r_1$ for each $\eta$ and the dotted vertical line indicated $d=\kappa$.} } \label{fig4}
\end{figure}

\textcolor{black}{
Lastly, we shift our focus to the $\phi$-layer dynamics, rewriting Eq.~(\ref{eq:02}) as $\dot{\phi}_i=\nu_i+\kappa\rho_1\sin(\varphi_1-\theta_i)+d\sin(\theta_i-\phi_1)$, where $\rho_1e^{i\varphi_1}=\frac{1}{N}\sum_{j=1}^Ne^{i\phi_j}$. The dynamics of each $\phi_i$ can be written in terms of a single sine term, however, due to the fact that all $\theta_i$'s are generically different, the target phase and amplitude are all distinct, making self-consistency analyses difficult. However, the microscopic properties may still be understood by comparing $\kappa$ and $d$. Namely, for $\kappa\gg d$ the dynamics are effectively undriven, in which case oscillators that become entrained form a single cluster. On the other hand, for $d\gg\kappa$ the driving term dominates, causing each $\phi_i$ to entrain with $\theta_i$, thereby mirroring any cluster behavior. This is illustrated in Fig.~\ref{fig4}, where we plot $\rho_1$ as a function of $d$ given $\kappa=3$ and $K=12$ for asymmetries $\eta=1$, $0.95$, $0.9$, $0.85$, and $0.8$ in blue, red, green, orange, and purple circles, respectively. Dashed horizontal lines denote the $r_1$ value for each $\eta$, illustrating that $\rho_1\to r_1$ in the $d\gg\kappa$ regime and in fact the 1-simplex layer can also store information. However, as $d$ decreases into the $d\ll\kappa$ regime this property is lost as the 1-simplex dynamics no longer reflect the structure in the 2-simplex layer.
}

\textcolor{black}{The analysis presented here} demonstrates how the extension from pair-wise to more general simplicial (specifically, $2$-simplex) interactions in coupled oscillator systems can give rise to a host of complex nonlinear phenomena \textcolor{black}{including information and memory storage}. Moreover, these phenomena can be captured and described using analytical methods. In particular, we have characterized a continuum of abrupt desynchronization transitions that occur at different coupling strengths without any abrupt synchronization transitions. This continuum stems from extensive multistability, whereby for sufficiently strong coupling an infinite number of partially synchronized states are stable. These different stable states represent synchronized states organized via different asymmetries into two clusters of entrained oscillators. In addition to highlighting the possible complexity that may arise in coupled oscillator systems with simplicial interactions, we hypothesize that simplicial interactions may give rise to novel nonlinear phenomena in other complex systems.

%Lastly, we note that similar, albeit more complicated dynamics and synchronization patterns emerge in $3$- and higher-order simplex interactions, that is, four-way interactions, five-way interactions, etc. Preliminary investigations (not shown here) indicate that in an $n$-simplex system oscillators organize into $n$ distinct, equidistant clusters around the circle. Abrupt desynchronization transitions persist, as does the extensive multistability, however the locations of transitions appears to increase with $n$ and the effect of asymmetry on the synchronized states themselves is more complicated. Other fruitful avenues of future research include the incorporation of nontrivial network structures and mixed coupling, i.e., the presence of both $n$- and $m$-simplex interactions with $n\ne m$.

\bibliographystyle{plain}

\end{document}